\documentstyle[12pt]{article}
\begin{document}
\setlength{\baselineskip}{0.30in}
\newcommand{\beq}{\begin{equation}}
\newcommand{\eeq}{\end{equation}}
\newcommand{\be}{\begin{equation}}
\newcommand{\ee}{\end{equation}}
\newcommand{\bi}{\bibitem}
\def\ne{\nu_e}
\def\nm{\nu_\mu}
\def\nt{\nu_\tau}

{\hbox to\hsize{November, 1997  \hfill TAC-1997-038}
\begin{center}
\vglue .06in
{\Large \bf { Varying leptonic chemical potentials and spatial variation of
primordial deuterium at high z.
  }
}
\bigskip
\\{\bf A.D. Dolgov}
 \\[.05in]
{\it{Teoretisk Astrofysik Center\\
 Juliane Maries Vej 30, DK-2100, Copenhagen, Denmark
\footnote{Also: ITEP, Bol. Cheremushkinskaya 25, Moscow 113259, Russia.}
}}\\
{\bf  B.E.J. Pagel }
 \\[.05in]
{\it{ NORDITA\\
Blegdamsvej 17,  DK-2100, Copenhagen, Denmark
 }}\\[.40in]

\end{center}
\begin{abstract}
We try to explain the spatial variation of primordial deuterium suggested by
some observations by varying leptonic chemical potentials. The variation of
the latter may take place in some scenarios of leptogenesis. The model predicts
a large mass fraction of $^4 He$ (35-60\%) and $^7 Li$ (up to  $10^{-9}$) in
deuterium-rich regions. Because of lepton family symmetry, the angular
variations of cosmic microwave background radiation can be sufficiently small
although still observable in future measurements.

\end{abstract}

Recently several groups [1-6]
have reported measurements of the deuterium abundance
in Lyman-limit absorption line systems with red-shifts $0.48<z<3.5$ on the
line of sight to quasars; these are believed to give essentially the
primordial value. Surprisingly some groups have claimed a high value,
$ D/H \approx 2\cdot 10^{-4}$ on the basis of ground-based data taken
with the Keck telescope, but this result is now thought to be due to various
errors \cite{tbk} and the best value available from two ``clean" systems
is $3\cdot 10^{-5}$ \cite{dt1}. However, Webb et al \cite{wcl} report a high
deuterium abundance,  $ D/H \approx 2\cdot 10^{-4}$, in an apparently clean
system with $z = 0.7$ observed with the Hubble Space Telescope, as well as
a low one in another system with $z = 0.5$, raising the possibility that
there might be real spatial variations in primordial $D/H$.

If the effect is indeed  real (which it is perhaps too early to judge),
its significance is difficult to overestimate.
It would
strongly change our approach to primordial nucleosynthesis and possibly to the
physics of the early universe. A possible variation of the light element
abundances was in fact considered in ref. \cite{dk} (see also \cite {ad1}),
where a model of leptogenesis was considered which, first, gave a large lepton
asymmetry, which could even be close to or larger than 1, and, second,
this asymmetry
might strongly change on astronomically large scales, $l_L$. The magnitude of
the latter depends on the unknown parameters of the model and can easily
be in the mega-giga parsec range. The model is based on the
Affleck-Dine~\cite{afd}
scenario of baryogenesis but in contrast to the original one it gives rise
to a large (and varying) lepton asymmetry and to a small baryonic one. Recently
a similar model of generation of large (but not varying) lepton asymmetry was
considered in ref. \cite {ccg}. In what follows we will not discuss the details
of the model but confine ourselves to a more phenomenological level, namely we
simply assume that there exists a mechanism which created large leptonic
asymmetries of order unity (electronic, muonic and/or tauonic) which vary
by 100\% over the distance $l_L$. A possible early universe scenario which
would give rise to such varying and large leptonic asymmetries will be
considered elsewhere. At the moment we put a less ambitious question:
whether it is
possible to describe the suspected spatial variation of primordial deuterium by
varying chemical potentials of neutrinos without conflict with the existing
astronomical data and what predictions can be made in such a model which can
be tested in future observations.

To explain the claimed variation of $^2 H$
at $z=0.7$ the magnitude of $l_L$ in terms of present-day units must be
smaller than or close to one gigaparsec. The lower bound on this scale $l_L$
may be much smaller. It can in principle be determined by measurements
of the abundances of light elements at large distances in our neighborhood,
say, $ z\geq 0.05$. It would be interesting if the scale $l_L$ coincides
with the 140/h Mpc scale observed in the large scale
structure of the universe \cite{lss,lss2}.

Another simple possibility to explain a varying abundance of deuterium is to
assume that the baryon-to-photon ratio in the universe varies as a function of
position. This idea was studied in ref. \cite{cos} where it was shown that the
necessary large scale isocurvature perturbations are excluded by the smallness
of angular fluctuations of the cosmic microwave background radiation (CMB).
A similar criticism is applicable at first sight to the model with varying
lepton asymmetry.
Indeed, it can be easily checked that the necessary change in chemical
potential
of electron neutrinos $\xi_{\nu_e}$ should be close to $-1$
to explain the possibly
observed variation of deuterium by roughly an order of magnitude. Such a change
in $\xi_{\nu_e}$ would induce a variation in total energy density during the RD
stage at a per cent level, which is excluded by the smoothness of CMB. However,
this objection can be avoided if there is a conspiracy between different
leptonic chemical potentials such that in different spatial regions they have
the same values but with interchange of electronic, muonic and/or tauonic
chemical
potentials. In other words we assume that in a particular spatial region the
three neutrino chemical potentials have the values
\beq{
 [\xi_{\nu_e},\,\xi_{\nu_\mu},\,\xi_{\nu_\tau}] =
[\alpha, \beta, \gamma] .
\label{alpha}
}\eeq
Then in another spatial region they
should have the same values but with an arbitrary interchange of $e$, $\mu$,
and $\tau$. This would ensure the same energy density
at different space points and small angular variations of CMB. In fact the
perturbations in CMB induced in this way would be non-vanishing and close to
existing observations. We will discuss them below. Since the abundances of
light elements are much more sensitive to the magnitude of the electron
neutrino
chemical potential than to those of
muon and tauon neutrinos, the variation of $\xi_{\nu_e}$
(accompanied by corresponding variations of $\xi_{\nu_\mu}$ and
$\xi_{\nu_\tau}$) would lead to a strong variation in the abundance of
deuterium and other light elements.

The equality of, say,  $\xi_{\nu_e}$ at one space point to $\xi_{\nu_\mu}$ at
another point looks like a very unnatural fine-tuning but this is not so.
The present theory of elementary particles is believed to be symmetric with
respect to interchange of three families of leptons. In the Affleck-Dine
type scenario of generation of charge asymmetry, the latter is generated
owing to the formation of baryonic (as in the original version) or leptonic
charge condensates along the so called flat direction in the potential of a
scalar field which possesses corresponding charges. It is rather natural to
assume that the potential respects the symmetry between different lepton
families. So if a flat direction corresponds to a scalar field with the
combination of leptonic charges  $[\alpha, \beta, \gamma]$, then there must
be flat directions with the same values of the leptonic charges but
interchanged
with respect to $e$, $\mu$, and $\tau$. In such a model there would be
regions with different values of leptonic chemical potentials which are
obtained
by transmutations of the original ones in (\ref{alpha}).

The symmetry between lepton families is broken at low energy by the masses of
charged leptons. So one may expect that there could be significant fluctuations
of the cosmic energy density when the temperature is close to the mass of
the charged tau-lepton ($m= 1777$ MeV) or to that of
the muon ($m = 106$ MeV).
We will see below that this is not the case. Another potential danger for a
model of this kind is the variation of the energy density associated with the
energy of the potential wall between the valleys (flat directions) with
different leptonic charges. These domain walls however disappear because all
scalar fields $\phi_l$ possessing different leptonic charges evolve down to the
same origin of the potential where they all vanish,  $\phi_l = 0$. Some
remnants
of the energy distortion remain but they have an energy density much smaller
than that of the original domain walls.

Let us turn now to the astro-phenomenology of the model concerning the
change in the light element abundances. It is straightforward and simple to
play with the standard nucleosynthesis code \cite{kaw} to study the influence
of different leptonic chemical potentials on the output of light elements
and in Table 1 we present a number of sample calculations from which we
can draw some combinations that would give a small amount of $^2 H$
in our neighborhood (and a larger one in deuterium-rich regions).
Considerations of Galactic chemical evolution \cite{pag} permit us to
infer that the abundance of primordial deuterium in nearby regions where
$^4 He$ is also measured is close to the low values determined at high
red-shift; we take as the best estimates for both these regions and
our neighborhood  $D/H = (3.2 \pm 0.8) \cdot 10^{-5}$
and $R(^4 He) = 0.24 \pm 0.01$, which are well fitted in the case of no
neutrino degeneracy for a baryon/photon ratio
$\eta_{10} = 5 \pm 1$. An adequate fit is also obtained for the same $\eta$ if
we take the combination $[\xi_{\nu_e}, \xi_{\nu_\mu}] = [0, -1]$, which gives
for the "mirror" region with $[\xi_{\nu_e}, \xi_{\nu_\mu}] = [-1, 0]$ a
substantially higher deuterium abundance $D/H = 8.5 \cdot 10^{-5}$. This
combination is not necessarily the best possible fit to the data,
be
the
%
but it seems too early to look for this, bearing
in mind that the observational data may change.
It is worth noting that if two (or all three)
$\xi$'s are permitted to vary, the
nucleosynthesis limits (for a recent reference see e.g. \cite{kks}) would
be invalidated.

One can see from Table 1 that the data somewhat resist the proposed
explanation. It would help if there is more deuterium in our neighborhood,
$\sim 5\cdot 10^{-5}$, and/or less in the deuterium-rich regions,
$\sim 10^{-4}$.
We did not try to use large values of chemical potentials because of
possible problems with smoothness of CMB temperature (see below). The
agreement with observations can be made better if all three chemical
potentials could be adjusted as free parameters.
In Table 2 we present the abundances of light elements for the {\it ad hoc}
choice $[\xi_{\ne},\xi_{\nm},\xi_{\nt}] = [-1,0.1,1]$ for $\eta_{10}=4$ and 5.
The last line may not be reliable because the program fails to converge.
It is noteworthy that  it is possible to have, besides high $^2 H$ and
$^4 He$ regions, regions with normal deuterium and low helium-4.

To describe simultaneously the suspected deuterium content
in the rich regions and that in our neighborhood
we need $\eta_{10} = 5-6$ and $\xi_{\ne} =0-0.1$ in the
poor regions and $\xi_{\ne} \approx -1.4$ in the rich regions. In this case
it is possible to get $D/H$ as large as $17\cdot 10^{-5}$ in rich regions.
The necessary value of $\xi$ is rather high and it would be easier for the
model if the deuterium fraction in rich regions would be around
$10\cdot 10^{-5}$.

A generic feature of our model is that simultaneously with high deuterium
a high mass fraction of helium-4 is predicted. It is at least 30-35\% or may be
even above 50\%. It is an interesting question what is the observational
upper bound on the abundance of $^4 He$ far away from us. All direct
measurements of $^4 He$ known to us were done at most at $z = 0.045$
corresponding  to a distance of $140 h^{-1}$ Mpc\cite{hel}. A very large
mass fraction of $^4 He$ can possibly be excluded with the help of star and
galaxy evolution. Stars should be brighter and have a shorter life-time. All
data indicate that distant objects (including quasars) have
more or less normal
chemical content. Still we do not know what is the permitted mass
fraction of helium-4 which does not contradict the data.  This
would be the  subject of a
separate study. Presumably 35\% of $^4 He$ in some
distant parts of the universe
is not excluded. As for much higher values, we do not have an answer now.

A very sensitive indicator of any inhomogeneities in the universe is the
cosmic microwave background. The bearers of electronic, muonic, and
tauonic chemical potentials have different masses: though different neutrinos
are most probably very light or even massless, so
that their contribution to the
energy density is the same, the masses of
the corresponding charged leptons are very
much different and this could be potentially dangerous for the model. This is
not the case, however, as can be seen from the following
considerations. Let us
assume for simplicity that there are only two lepton families, electronic and
muonic. Let us assume also that the primeval plasma has nonzero electronic
and muonic charge densities, $D_e$ and $D_\mu$. Of course the plasma is
electrically neutral. Thermal equilibrium in the plasma is fulfilled with
a very good accuracy (at least for temperatures above 3 MeV, when
neutrinos decouple). The distributions of different particles are given by
the normal Fermi (or Bose) functions with nonzero chemical potentials which
permit to have nonzero $D_e$ and $D_\mu$. Due to reactions
$ e^- +\bar \nu_e \leftrightarrow \mu^- + \bar \nu_\mu$ and similar (crossed)
ones, the following relation between chemical potentials must be fulfilled in
thermal equilibrium:
\beq
\xi_e - \xi_{\nu_e} = \xi_\mu - \xi_{\nu_\mu}
\label{xie}
\eeq
There is also the condition of electric neutrality of the plasma:
\beq
\delta n_e + \delta n_\mu =0
\label{qel}
\eeq
and the expressions for electronic and muonic charge densities:
\beq
\delta n_e + \delta n_{\nu_e} = D_e
\label{de}
\eeq
and
\beq
\delta n_\mu + \delta n_{\nu_\mu} = D_\mu
\label{dmu}
\eeq
where $\delta n_a = n_a - n_{\bar a}$ is the difference in number densities
of particles and antiparticles with
\beq
 n_a =  \int {d^3 p \over 1 + \exp (E/T - \xi_a )}
\label{na}
\eeq
The number density of antiparticles is given by the same expression with the
opposite sign of $\xi_a$.

One can see from the symmetry property of the system of equations
(\ref{xie}-\ref{dmu}) that for any solution corresponding to the set
$[D_e, D_\mu]=[\alpha,\beta]$ there exists the mirror solution corresponding
to the set $[D_e, D_\mu]=[\beta, \alpha ]$ which can be constructed from the
original solution by the substitution:
$\delta n_e\leftrightarrow -\delta n_e$,
$\delta n_\mu\leftrightarrow -\delta n_\mu$ (correspondingly
$\xi_{e,\mu}\leftrightarrow -\xi_{e,\mu}$) and
$\delta n_{\nu_e}\leftrightarrow \delta n_{\nu_\mu}$.
Evidently the energy densities of both
solutions are the same.

There are some other possible inhomogeneities in the energy density that
could be either dangerous for the model or observable in CMB.
The first and most evident one is related to the binding
energy of $^4 He$, which is 7 MeV per nucleon. Since the mass fraction
of $^4 He$ may change by a factor of 2 in deuterium- (and helium-) rich regions
(from 25\% to more than 50\%),
this means that the variation in baryonic energy density may be as large as
$2\cdot 10^{-3}$. The contribution of baryons to the total energy density
is given by $\Omega_B = 3\% (\eta/4) (0.65/h)^2$, so the relative
density fluctuations are at most
\be
{\delta \rho \over \rho_{tot} } = 6\cdot 10^{-5} (\eta /4) (0.65/h)^2
\label{drho}
\ee
To estimate the fluctuations in CMB temperature we can use the results of
ref. \cite{cos}, where similar isocurvature density perturbations, but
with amplitude $(2\cdot 10^{-3})^{-1} =500\times$ larger, were considered.
According to their results normalized to our smaller perturbations
\be
{\delta T \over T} = 10^{-5} \left( {\lambda_0 \over 10\lambda}\right)^2
\label{dtt}
\ee
where $\lambda_0 = c/H_0 = 3 {\rm Gpc} /h$. So the anisotropy induced by
the Sachs-Wolfe effect would be below the observational bounds for
scales above $\sim 300 h^{-1}$ Mpc. If D/H in the rich regions is about
$10^{-4}$ (and not $2\cdot 10^{-4}$), then the variation of helium-4 could
be smaller. Correspondingly smaller density perturbations would be induced.
In this case  smaller scales in fluctuations of CMB temperature, down to
$\sim 150 h^{-1}$ Mpc, would be permitted. However, on scales below $2^o$
(or below 200 Mpc), the result (\ref{dtt}) would not be valid. These scales
are dominated by Doppler shift across the fluctuations at the surface
of last scattering \cite{js}. The measurements on small scales permit
possibly $\delta T /T = 3\cdot 10^{-5}$, which could also be compatible with
this model. Such fluctuations may be observed in the future MAP or
PLANCK missions or with balloons.

There is another effect which is more subtle theoretically
but which could also give rise to similar fluctuations in
$\delta T /T$. The energy densities of electron and muon neutrinos
are known \cite{df,dt} to be different owing to the following effect. After
neutrinos decoupled from the primeval plasma, which roughly took place at
$T = 2$ MeV for electron neutrinos and at $T = 3$ MeV for muonic and tauonic
ones, the temperatures of electrons and photons became somewhat different from
the neutrino temperature owing to heating of the electromagnetic component of
the plasma by $e^+ e^-$-annihilation into photons. Because of this temperature
difference and due to residual $e^+ e^-$-annihilation into $\nu \bar \nu$, the
usually assumed equilibrium neutrino distributions became slightly
distorted. The nonequilibrium correction to the
energy density of electron neutrinos in the standard model
is approximately \cite{dhs}:
\be
\Delta \rho_{\ne} / \rho_\nu \approx 0.9\%
\label{drhoe}
\ee
and the distortion of the energy density of muon and tauon neutrinos
is
\be
\Delta \rho_{\nm} / \rho_\nu =\Delta \rho_{\nt} / \rho_\nu\approx 0.4\%
\label{drhom}
\ee
(closely similar results are obtained in ref. \cite{hm}).
The difference between $\nu_e$ and $\nu_{\mu,\tau}$ is related to
a greater efficiency of the process $e^+ e^- \rightarrow \nu \bar \nu$ due
to the presence of charged current interactions only for $\nu_e$. Now
because
of nonzero leptonic chemical potentials these results
would slightly change. They
would remain the same in the Boltzmann approximation because the probability
of $e^+ e^-$-annihilation into $\nu \bar \nu$ does not depend on the chemical
potential of neutrinos in the case of Boltzmann statistics. Typically
corrections due to Fermi statistics are about 10\%. So the relative
efficiency of annihilation due to nonzero chemical potentials becomes smaller
by approximately $0.1 [ \cosh (\xi) -1]$. This expression is true for
relatively small $\xi$, $\xi \leq 1$; for larger $\xi$ it is changed to
a power law.

To get an estimate of the magnitude of the
density fluctuations due to variation of chemical potentials let us assume
that in our neighborhood the chemical potentials have
the values $\xi_{\ne} =\xi_{\nt} = 0$ and $\xi_{\nm} = -1$ and in the
deuterium-rich
region they are $\xi_{\nm} =\xi_{\nt} = 0$ and $\xi_{\ne} = -1$.
Thus the relative energy density of neutrinos changes by
\be
{\delta \rho_\nu^{(tot)} \over  \rho_\nu }=
\delta \left({\Delta \rho_{\ne} \over \rho_\nu }
+ {\Delta \rho_{\nm} \over \rho_\nu }+
{\Delta \rho_{\nt} \over \rho_\nu } \right) =
(0.9\%-0.4\%) \cdot 0.1 (\cosh \xi -1) \sim 2.5 \cdot 10^{-4}
\label{drtot}
\ee
Keeping in mind that one
neutrino species contributes 10-20 \% to the total energy density
during the RD stage, we find that the relative density fluctuations
of neutrinos due to variation of chemical
potentials are approximately
$\delta \rho_\nu /\rho_{tot} \approx 5\cdot 10^{-5}$.
In fact the fluctuations of the total energy density are very much smaller
than that because the increase in $\rho_\nu$ is accompanied by a similar
decrease in the energy density of photons and $e^{\pm}$. Thus the phenomenon
we discuss gives rise to a rather peculiar perturbation: the variation of the
total energy density  is negligibly small but the radiation temperature
varies between different spatial points.

As was calculated in  ref. \cite{dhs} the photon temperature drops in
comparison with the standard one by $10^{-3}$, due to the
above mentioned transfer of energy from
the electromagnetic component of the
plasma to neutrinos. This change of temperature should be proportional
in a crude approximation to the above mentioned change of neutrino energy:
\be
{\Delta T \over T} \approx
0.1 {\Delta \rho_\nu ^{(tot)} \over \rho_\nu} \sim 10^{-3}
\label{dtt2}
\ee
Now if chemical potentials are not spatially constant, the quantity
$ \Delta \rho_\nu^{(tot)} /\rho_\nu $ would vary together with the chemical
potentials at different points. Its variation is given by eq. (\ref{drtot}).
Accordingly the variation of the photon temperature due to this effect is
\be
{\delta T \over T } = \delta \left( 0.1 {\Delta \rho_\nu^{(tot)}
\over \rho_\nu }\right) \approx 5\cdot 10^{-5} (\cosh \xi -1)
\sim 2.5 \cdot 10^{-5}
\label{dtt3}
\ee
which is close to the observational bounds.

These are of course very crude estimates. The real result should be somewhat
smaller. An account of inverse annihilation,
$\bar \ne \ne \rightarrow e^- e^+ $ and of elastic $\nu e$-scattering
results in a smoothing down of the spectral distortion. An estimate of these
"inverse" effects, made along the lines of semi-analytical estimates of
ref. \cite{df}, diminishes the temperature change by a
factor of roughly $2/3$.

The magnitude of temperature fluctuations depends in particular on the
unknown values of the $\xi$'s. For example for $|\xi| = 0.7$ the effect
would be twice smaller than for $|\xi| = 1$, while for $|\xi | =1.4$ it
is twice bigger. We take $|\xi | =1.4$ as an upper limit for the magnitude of
possible variations of chemical potentials. To be on the safe side we
possibly need somewhat smaller $\xi$'s and correspondingly a fraction of
deuterium in the rich regions of about $10^{-4}$.
A rigorous calculation of
the effect is a straightforward but  formidable numerical problem. It seems
premature to do that at this stage.  However, we will have to do the
calculations if the effect of spatial variation in deuterium abundances
is confirmed and the predicted variation of helium-4 is either found or
not ruled out.

To conclude, we try to explain a spatial variation of primordial
deuterium, that has perhaps been observed,
by varying leptonic chemical potentials. The model could be confirmed
(or rejected) by looking for a very large mass fraction of primordial
helium-4 in deuterium-rich regions but this is not a practical possibility
in the context of data available now or in the near future.
A more promising test seems to be possible from the theory of stellar evolution
with a high mass fraction of $^4 He$.
The hypothesis also predicts  larger abundances of
other light elements in these regions, e.g. $^7 Li$ should be at the level of
$10^{-9}$. There might be also regions with normal deuterium and low
helium-4.
If there is a family symmetry which ensures permutational symmetry
for different chemical potentials, a very large distortion of CMB isotropy
can be avoided, but there still remain nonzero
$\delta T /T$ fluctuations  which can be
detected if the higher abundances of deuterium and other elements
(in particular $^4 He$) exist. It is a curious coincidence that the
theory of large scale structure formation may possibly favor neutrino
chemical potentials \cite{lm} close to those that are needed in our model.

\bigskip
{\bf Acknowledgments}
The work of A.D. was supported by Danmarks Grundforskningsfond through its
funding of the Theoretical Astrophysical Center. We thank I. Novikov and
J. Madsen for stimulating comments.

\newpage
\begin{table}
\begin{center}
\begin{tabular}{|c|| r |r| c|| c| c| c|}
\hline
&&&&&& \\
$ \eta_{10}$ &~~$\xi_{\ne}$~~&~~$\xi_{\nm}$~~&~~$\xi_{\nt}$
~&~$10^5\, {D\over H}$~&~$Y_p$~~&~~$10^{10}\, {^7 Li\over H}$~~\\
&&&&&&\\
\hline
 4 & 0 & 0 & 0 & 5.03 & 0.242 & 1.85 \\
& 0 & $-$1 & 0 & 7.33 & 0.248 & 1.78 \\
& $-$1 & 0 & 0 & 12.1 & 0.539 & 4.45 \\
 & 0 &$-$1.3  & 0 & 5.56  & 0.252 & 1.74 \\
 & $-$1.3 & 0 & 0 & 20.0 & 0.644 & 10.6 \\
      & 0.1 & $-$1 & 0 & 5.07 & 0.224 & 1.66 \\
      & $-$1 & 0.1 & 0 & 12.1 & 0.539 & 4.45 \\
 & 0.1 & $-$1.3 & 0 & 5.27 & 0.228 & 1.62 \\
 & $-$1.3 & 0.1 & 0 & 20.0 & 0.644 & 10.6 \\
\hline
 5 & 0 & 0 & 0 & 3.55  & 0.244 & 2.95 \\
   & 0 & $-$1 & 0 & 3.76 & 0.250 & 2.82 \\
   & $-$1 & 0 & 0 & 8.50 & 0.544 & 4.40 \\
       & 0 & $-$1.4 & 0 & 3.98 & 0.256 & 2.70 \\
       & $-$1.4 & 0 & 0 & 17.1 & 0.686 & 10.9 \\
& 0.1 & $-$1 & 0 & 3.57 & 0.226 & 2.64 \\
& $-$1 & 0.1 & 0 & 8.51 & 0.544 & 4.40 \\
    & 0.1 & $-$1.4 & 0 & 3.78 & 0.231 & 2.53 \\
    & $-$1.4 & 0.1 & 0 & 17.2 & 0.686 & 10.9 \\
\hline
6 & 0 & 0 & 0 & 2.65  & 0.246 & 4.35 \\
   & 0 & $-$1 & 0 & 2.82 & 0.252 & 4.17 \\
   & $-$1 & 0 & 0 & 6.40 & 0.548 & 5.36 \\
       & 0 & $-$1.4 & 0 & 2.98 & 0.258 & 3.99 \\
       & $-$1.4 & 0 & 0 & 12.8 & 0.692 & 8.83 \\
& 0.1 & $-$1 & 0 & 2.67 & 0.228 & 3.90 \\
& $-$1 & 0.1 & 0 & 6.41 & 0.548 & 5.36 \\
    & 0.1 & $-$1.4 & 0 & 2.83 & 0.233 & 3.74 \\
    & $-$1.4 & 0.1 & 0 & 12.8 & 0.692 & 8.84 \\
\hline

\end{tabular}
\bigskip
\caption{Abundances of light elements for different values of the baryon
number density, $\eta = 10^{10} n_B / n_\gamma$ and neutrino chemical
potentials $\xi_{\nu_a}$.}

\end{center}
\end{table}

\newpage
\begin{table}
\begin{center}
\begin{tabular}{|c|| r |r| r|| c| c| c|}
\hline
&&&&&& \\
$ \eta_{10}$ & $\xi_{\ne}$~~&~~$\xi_{\nm}$~~&~~$\xi_{\nt}$
~&~$10^5\, {D\over H}$~&~$Y_p$~~&~~$10^{10}\, {^7 Li\over H}$~~\\
&&&&&&\\
 \hline
 4&0.1 & $-$1 & 1 & 5.35 & 0.229 & 1.61 \\
    &  $-$1 & 0.1 & 1 & 13.2 & 0.548 & 4.84 \\
 &  1 &$-$1  & 0.1 & 3.98  & 0.080 & 0.70 \\
\hline
 5&0.1 & $-$1 & 1 & 3.77& 0.231 & 2.54 \\
    &  $-$1 & 0.1 & 1 & 9.21 & 0.553 & 4.49 \\
 &  1 &$-$1  & 0.1 & 2.80  & 0.081 & 1.12 \\
 \hline

\end{tabular}
\bigskip
\caption{Abundances of light elements for
  $\eta = 10^{10} n_B / n_\gamma=4,\,\,5$ and nonzero values of all three
neutrino chemical potentials $\xi_{\nu_a}$.}

\end{center}
\end{table}

\newpage

\end{document}